\documentclass{emulateapj}
\begin{document}

\def\bullet{\object{1E0657$-$56}}
\def\bbullet{\object{MACS~J0025.4$-$1222}}

\def\arcsecf{\!\!^{\prime\prime}}
\def\arcminf{\!\!^{\prime}}
\def\diff{\mathrm{d}}
\def\ngx{N_{\mathrm{x}}}
\def\ngy{N_{\mathrm{y}}}
\def\eck#1{\left\lbrack #1 \right\rbrack}
\def\eckk#1{\bigl[ #1 \bigr]}
\def\rund#1{\left( #1 \right)}
\def\abs#1{\left\vert #1 \right\vert}
\def\wave#1{\left\lbrace #1 \right\rbrace}
\def\ave#1{\left\langle #1 \right\rangle}
\def\kms{{\rm \:km\:s}^{-1}}
\def\dds{D_{\mathrm{ds}}}
\def\dd{D_{\mathrm{d}}}
\def\ds{D_{\mathrm{s}}}
\def\cs{\mbox{cm}^2\mbox{g}^{-1}}
%===============================================================================

\title{Revealing the properties of dark matter in the merging cluster {\bbullet}\altaffilmark{*}} \altaffiltext{*}{Based on
observations made with the NASA/ESA Hubble Space Telescope, obtained
at the Space Telescope Science Institute, which is operated by the
Association of Universities for Research in Astronomy, Inc., under
NASA contract NAS 5-26555. These observations are associated with
program \# GO11100 and GO10703.  This work is also based on
observations collected at the W.M. Keck Observatory. }
\shorttitle{}
\author{Maru\v{s}a Brada\v{c}\altaffilmark{1,x},
Steven W. Allen\altaffilmark{2},
Tommaso Treu\altaffilmark{1,y},
Harald Ebeling\altaffilmark{3},
Richard Massey\altaffilmark{4},
R.\ Glenn Morris\altaffilmark{2}, 
Anja von der Linden\altaffilmark{2},
Douglas Applegate\altaffilmark{2},
}
\shortauthors{Brada\v{c} et al.}
\altaffiltext{1}{Department of Physics, University of California, Santa Barbara, CA 93106, USA}
\altaffiltext{2}{Kavli Institute for Particle Astrophysics and Cosmology,
Stanford University, 382 Via Pueblo Mall, Stanford, CA 94305-4060, USA}
\altaffiltext{3}{Institute for Astronomy, University of Hawaii, 2680 Woodlawn Drive, Honolulu, HI 96822, USA}
\altaffiltext{4}{Institute for Astronomy, Royal Observatory Edinburgh, Blackford Hill, Edinburgh EH9 3HJ, UK}
\altaffiltext{x}{Hubble Fellow}
\altaffiltext{y}{Sloan Fellow, Packard Fellow}
\email{marusa@physics.ucsb.edu}

%-------------------------------------------------------------------------------

\begin{abstract}
We constrain the physical nature of dark matter using the newly
identified massive merging galaxy cluster {\bbullet}. As was
previously shown by the example of the Bullet Cluster (1E0657-56),
such systems are ideal laboratories for detecting isolated dark
matter, and distinguishing between cold dark matter (CDM) and other
scenarios (e.g.\ self-interacting dark matter, alternative gravity
theories). {\bbullet} consists of two merging subclusters of similar
richness at $z=0.586$.  We measure the distribution of X-ray emitting
gas from Chandra X-ray data and find it to be clearly displaced from
the distribution of galaxies. A strong (information from highly
distorted arcs) and weak (using weakly distorted background galaxies)
gravitational lensing analysis based on Hubble Space Telescope
observations and Keck arc spectroscopy confirms that the subclusters
have near-equal mass. The total mass distribution in each of the
subclusters is clearly offset (at $>4\sigma$ significance) from the
peak of the hot X-ray emitting gas (the main baryonic component), but
aligned with the distribution of galaxies. We measure the fractions of
mass in hot gas ($0.09^{+0.07}_{-0.03}$) and stars ($0.010^{+0.007}_{-0.004}$),
consistent with those of typical clusters, finding that dark matter is
the dominant contributor to the gravitational field.  Under the
assumption that the subclusters experienced a head-on collision in the
plane of the sky, we obtain an order-of-magnitude estimate of the dark
matter self-interaction cross-section of $\sigma/m < 4\cs$,
re-affirming the results from the Bullet Cluster on the collisionless
nature of dark matter.
\end{abstract}

%-------------------------------------------------------------------------------

\section{Introduction}
\label{sec:intro}
One of the most important quests in cosmology is to understand the
formation and evolution of galaxies and galaxy clusters. While the
standard paradigm of cold dark matter (CDM) within a $\Lambda$CDM
cosmology works well overall, discrepancies between the predictions of
this model and observations remain (such as the abundance of
substructure in halos, and the shapes of dark matter profiles,
especially in dwarf systems). These issues can be resolved in several
ways, for example by invoking warm and/or self-interacting dark
matter.

Mergers of galaxy clusters are excellent laboratories to test such
ideas. Not only are cluster mergers the most energetic events since
the Big Bang, but, as demonstrated by a recent study of the Bullet
Cluster (1E0657$-$56; \citealp{clowe04,bradac06,clowe06}), they
provide a unique avenue by which to probe the self-interaction
cross-section of dark matter particles and test alternative gravity
theories.

In clusters with recent major merger activity, the positions of the
dark matter and main baryonic component (the X-ray emitting gas) can
become temporarily separated.  This occurs because the gas is
collisional and experiences ram pressure, whereas galaxies and
(presumably) dark matter are effectively collisionless. A merging
direction perpendicular to the line-of-sight maximizes the apparent
separation, allowing observers to independently study the main
baryonic and dark matter components. In the example of the Bullet
Cluster, \citet{clowe04},
\citet{bradac06}, and \citet{clowe06} used X-ray and gravitational
lensing data to derive mass maps for both the baryonic and dark matter
components, clearly showing the presence and dominance of dark matter.
In agreement with the standard CDM paradigm, these observations
indicate that dark matter has to be in a near-collisionless form, with
a scattering cross-section $\sigma/m < 0.7\cs$ \citep{randall07}.
 
A recent study of the merging cluster A520 by \citet{mahdavi07} rests
less easily within this simple picture. For A520, \citet{mahdavi07}
find a central mass peak that contains gas and dark matter but no
significant galaxy concentration. These authors argue that such a
configuration could occur if dark matter is collisional with a
self-interaction cross-section of $3.8 \pm 1.1\cs$, a result that is
inconsistent with the Bullet Cluster limits.  However, the statistical
significance of the result for A520 is not high and the dynamics of
the cluster are more complicated than {\bullet}, with A520 appearing
to be in the late stages of multiple, complex merger events.

It has also been argued that the optical and X-ray
observations of {\bullet} might be explained without standard CDM, by instead
modifying the theory of gravity (and thus the effects of gravitational
lensing). Models using the modified Newtonian Dynamics (MOND) theory
\citep[e.g.][]{milgrom83,bekenstein04} have, however, so far achieved
only limited success in reproducing the weak lensing observations
(e.g.\ \citealp{angus06, brownstein07}) and are unable to explain the
observed strong lensing. Moreover, the application of MOND models to
galaxy clusters in general still requires the presence of large masses
of invisible matter, which are typically argued to be a mixture of
massive neutrinos and cold baryons \citep[e.g.][]{angus07}.

Clearly, a satisfactory resolution to this debate requires further
observations of massive, merging clusters; preferably those with
simple, two-body configurations. Using deep, ground-based optical
imaging and short, snapshot Chandra exposures of clusters in the
all-sky, X-ray flux-limited MAssive Cluster Survey (MACS)
\citep{ebeling01, ebeling07}, we have identified the X-ray brightest,
most massive, obvious post-merging clusters in the redshift range
$0.3<z<0.7$. From this search, {\bbullet} emerges as a massive,
merging cluster with an apparently simple geometry, consisting of two
large subclusters of similar richness, both at redshift $z=0.586$,
colliding in approximately the plane of the sky.

Owing to their high masses and favorable redshifts, MACS clusters are
ideally suited to gravitational lensing studies. Here we use new,
multi-color Hubble Space Telescope (HST) data to determine a projected
mass map for {\bbullet}, using both strong and weak lensing
information. When combined with the Chandra data, this allows us to
map the distributions of dark matter, hot X-ray emitting gas, and
stars. The dark matter is resolved into two main clumps, centered on
the galaxy populations.  The Chandra data reveal a single, clear X-ray
peak located between the two optical/lensing subclusters. The observed
offset between the locations of the hot gas and galaxies/lensing mass
is in good agreement with the expectations for standard collisionless
cold dark matter.  Both {\bullet} and {\bbullet} provide direct
evidence for the existence of dark matter and can be used to estimate
the self-interaction cross-section.

This paper is structured as follows. Section~\ref{sec:data} describes
the optical and X-ray data and reduction procedures. Section 3
discusses the dynamics of the cluster, based on optical spectroscopy.
Section~\ref{sec:swdata} presents the strong and weak gravitational
lensing analysis and the mass reconstruction based on these
data. Section~\ref{sect:xray_analysis} discusses the X-ray data
analysis. Section~\ref{sec:stellarmass} details the results on the
stellar and X-ray gas mass fractions. 
In Section~\ref{sec:bardm} we present our main
result on the relative distributions of dark and baryonic matter in
the cluster, and determine rough limits on the dark matter
self-interaction cross-section. Our conclusions are summarized in
Section~\ref{sec:conclusions}.

Throughout the paper we assume a $\Lambda$CDM cosmology with
$\Omega_{\rm m} = 0.3$, $\Omega_{\Lambda} = 0.7$, and Hubble constant
$H_0 = 70 {\rm \ km \: s^{-1}\:\mbox{Mpc}^{-1}}$. At a redshift
$z=0.586$, the length scale is 6.61 kpc/arcsec.

\section{Observations of \protect\bbullet}
\label{sec:data}

We observed {\bbullet} with HST/ACS for $4140\mbox{s}$ in the F555W
filter and $4200\mbox{s}$ in F814W (6 exposures each) as part of the
Cycle 13 proposal GO-10703 (PI Ebeling, November 5 2006). We also
obtained a $11000\mbox{s}$ F450W HST/WFPC2 observation (11 exposures)
as a part of the Cycle 16 proposal GO-11100 (PI Brada\v{c}, June 9 2007).  As a
part of the MACS survey, the cluster has been observed in the BVRIz'
bands with Subaru Suprime-Cam.  Here we use ACS F814W as our primary
weak lensing band because it has the largest number of resolved
background sources for which shapes can be measured, and a well
understood point spread function (PSF)
\citep{rhodes07}. The other bands are used to aid the color matching
for identifying strongly lensed systems and to exclude foreground
objects from the weak lensing catalogs.

The demands placed by the lensing analysis require special care when
reducing the images. The ACS images were reduced using the HAGGLES
pipeline (P. Marshall et al. 2008, in preparation).  A similar
procedure was also followed for the WFPC2 data. The data reduction
procedure is described in \citet{bradac08}.

A sample of more than $200$ objects in {\bbullet} were targeted in the
course of a DEIMOS/KECK spectroscopic campaign for the MACS survey
(\citealp{ebeling07}). The DEIMOS data were reduced using the same procedure as described in \citet{ma08}. In addition, LRIS/KECK
\citep{LRIS} has been used to target likely multiply-imaged sources
selected based on their colors in the HST data. One slitmask was used
with an exposure of $6.3\mbox{ks}$ on September 13 2007. The LRIS data
were reduced using a dedicated pipeline, developed and kindly made
available by M.\ Auger (see also \citealp{auger08}).

{\bbullet} has been observed twice with the Chandra X-ray Observatory,
in November 2002 for 19.3 ksec (obsid 3251); and in August 2004 for
24.8 ksec (obsid 5010).  Both observations were made with ACIS-I in
VFAINT mode. We have analysed the data using version 3.4 of the CIAO
software and CALDB version 3.3.0.1. Applying conservative filtering,
we obtain clean exposures of 15.7 and 21.3 ksec for obsid 3251 and
5010, respectively (total 37 ksec).  These two exposures were combined
for imaging purposes, but fitted separately and simultaneously in
the spectral analysis.

\section{Optical dynamical analysis}
\label{sec:dyn}

\begin{figure}[ht!]
\begin{center}
\includegraphics[width=0.4\textwidth]{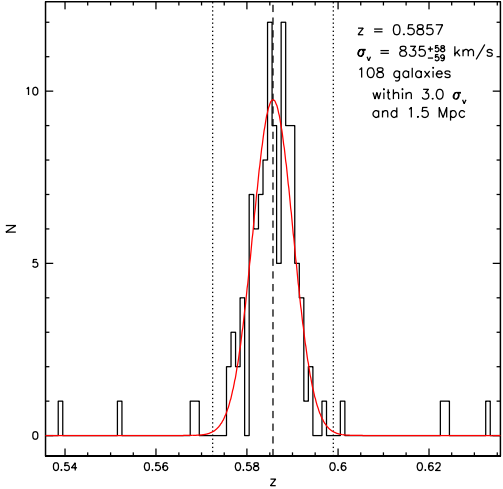}
\end{center}
\caption{Histogram of the redshift distribution around the cluster
  redshift. Only galaxies within projected 1.5~Mpc are shown. The
  dashed line indicates the cluster redshift, the dotted lines the
  $\pm3\sigma_{\rm Cl}$ limits used to calculate the cluster redshift
  and velocity dispersion. The Gaussian curve overplotted in red
  illustrates the measured velocity dispersion.}
\label{fig:specmemb}
\end{figure}

Using the redshifts obtained with DEIMOS, we measure the cluster
redshift $z_{\rm Cl}$ and velocity dispersion $\sigma_{\rm Cl}$. This
is done via an iterative algorithm to determine $z_{\rm Cl}$ and
$\sigma_{\rm Cl}$ from galaxies within 1.5~Mpc from the peak of the
X-ray emission (\S\ref{sect:xray_analysis}), and within
$\pm3\sigma_{\rm Cl}$ from $z_{\rm Cl}$. Both $z_{\rm Cl}$ and
$\sigma_{\rm Cl}$ are calculated using the biweight estimator of
\citet{bfg90}. Galaxies are allowed to re-enter the sample at each
iteration. This yields $z_{\rm Cl}=0.5857$ and $\sigma_{\rm Cl} =
835^{+58}_{-59}\kms$, measured from 108 galaxies. The overall
appearance of the redshift distribution is roughly Gaussian, with
little evidence for a bimodal distribution (Fig.~\ref{fig:specmemb}).

As a test for the presence of dynamical substructure, we compute the
statistics suggested by \citet{drs88}. Briefly, this method measures
local deviations from the overall cluster redshift and velocity
dispersion, and compares the result to Monte Carlo realizations of the
redshift and position values, where the redshifts have been
randomized. The probability of finding larger deviations in the
randomized cluster realizations is 3\%, i.e. cluster substructure is
marginally detected. This is mainly driven by some of the galaxies in
the north-western subcluster, which have slightly lower than systemic
redshifts.

  The south-eastern (SE) subcluster has two equally-dominant galaxies (BCG1
and BCG2; their magnitudes are equal within the errorbars; see
Fig.~\ref{fig:arcs}) of redshifts $z=0.5856 \pm 0.0003$ and $z=0.5838
\pm 0.0003$. The north-western (NW) one has a single dominant galaxy
(BCG3) at a redshift $z=0.5842 \pm 0.0003$. Assuming that the dominant
galaxies lie at or near the minimima of the subcluster potentials, and
averaging the redshifts of BCG1 and BCG2 in the SE subcluster, we
infer a redshift difference between the NW and SE clumps of $\Delta z=0.0005
\pm 0.0004$.  This implies a line of sight velocity difference between
the subclusters of $100\pm80\kms$.  Given that the subclusters have
collided, they have likely decoupled from the Hubble flow.

Based on this result and the measured velocity dispersion, we can
obtain a crude estimate of the merger velocity and the angle of the
merger axis relative to the plane of the sky. For {\bullet},
\citet{barrena02} measure a velocity dispersion of
$1132^{+117}_{-106}$ km/s (from 52 early-type galaxy spectra) and
\citet{markevitch06} estimate a merger velocity for {\bullet} (based
on a 500ks Chandra observation) of $\sim 4500$ km/s (although this
value is thought to be extreme;
e.g. \citealp{hayashi06,springel07}). Scaling by the ratio of these
velocity dispersions, and assuming that the merger velocity in
{\bbullet} is more typical for a cluster of its mass, we crudely
estimate a merger velocity of $\sim2000\kms$. Given this merger
velocity, the strong evidence that the subclusters in {\bbullet} have
merged recently, and the small, observed line-of-sight velocity
difference between the subclusters, the merger event is likely to have
occured in close to the plane of the sky (within 5 degrees). The
separation between the subclusters implies that closest approach
occured a few $10^8$ years ago. Deeper X-ray data are required to refine
our understanding of the geometry of the cluster.

\section{Lensing Analysis}
\label{sec:swdata}

The optical imaging data were used to identify multiply imaged sources (for strong lensing) and measure the distortion of background galaxies (for weak lensing). We then simultaneously fit these two data sets to obtain the (total) mass distribution of {\bbullet}.

\subsection{Strong lensing features}

\begin{figure*}[ht]
\begin{center}
\begin{tabular}{ccc}
\multicolumn{3}{c}{\includegraphics[width=0.5\textwidth]{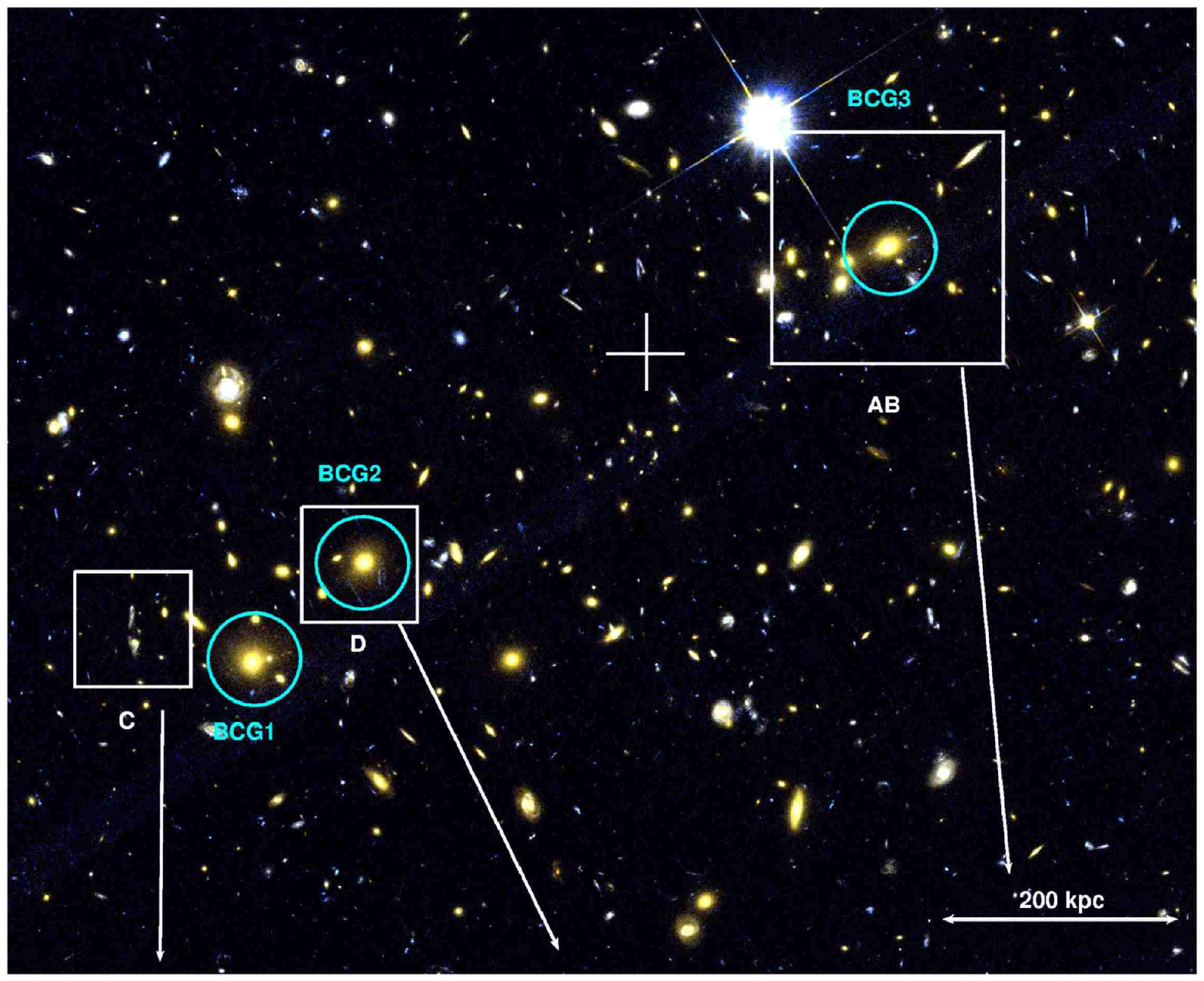}} \\
\includegraphics[width=0.33\textwidth]{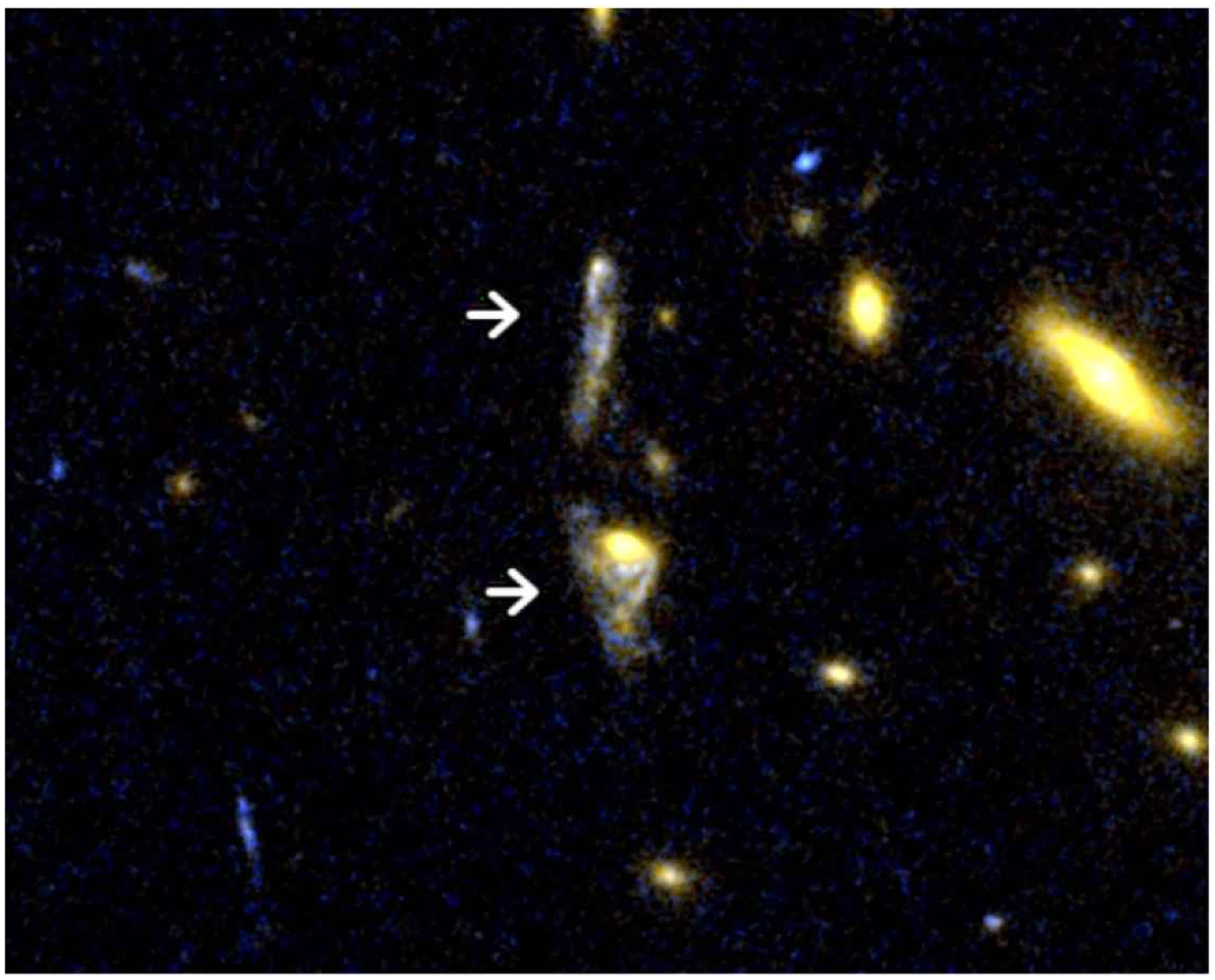} &
\includegraphics[width=0.33\textwidth]{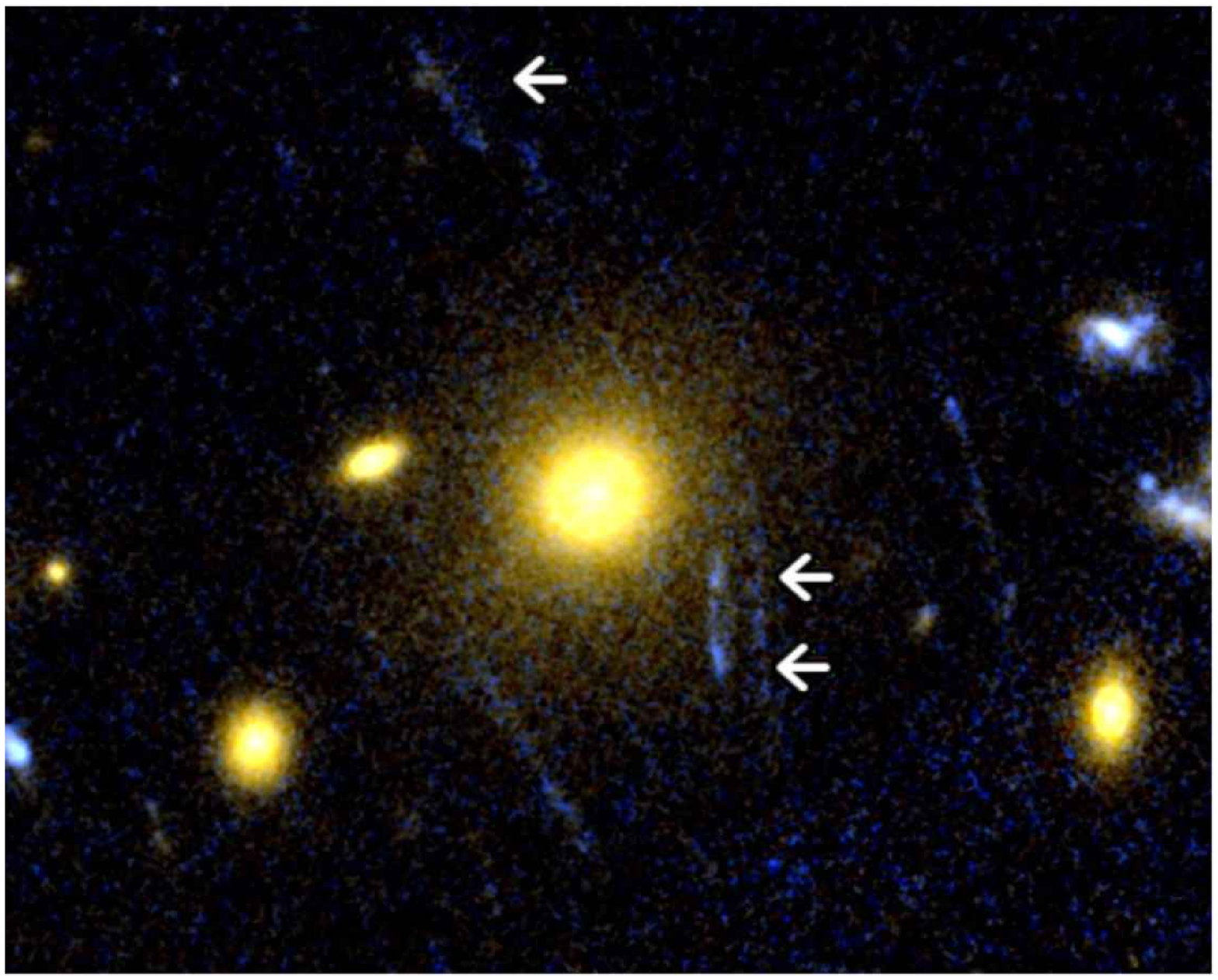} &
\includegraphics[width=0.33\textwidth]{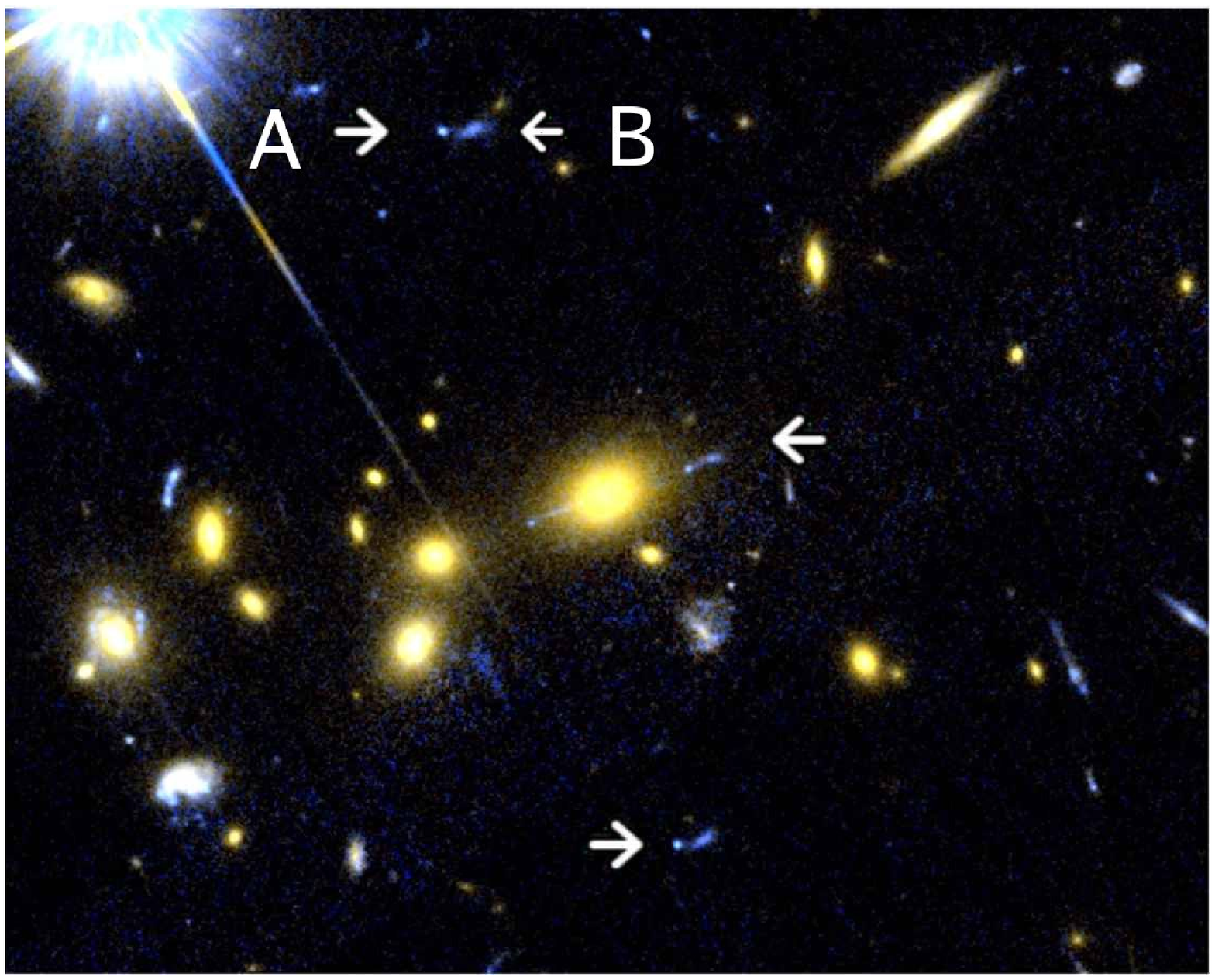} \\
\includegraphics[width=0.33\textwidth]{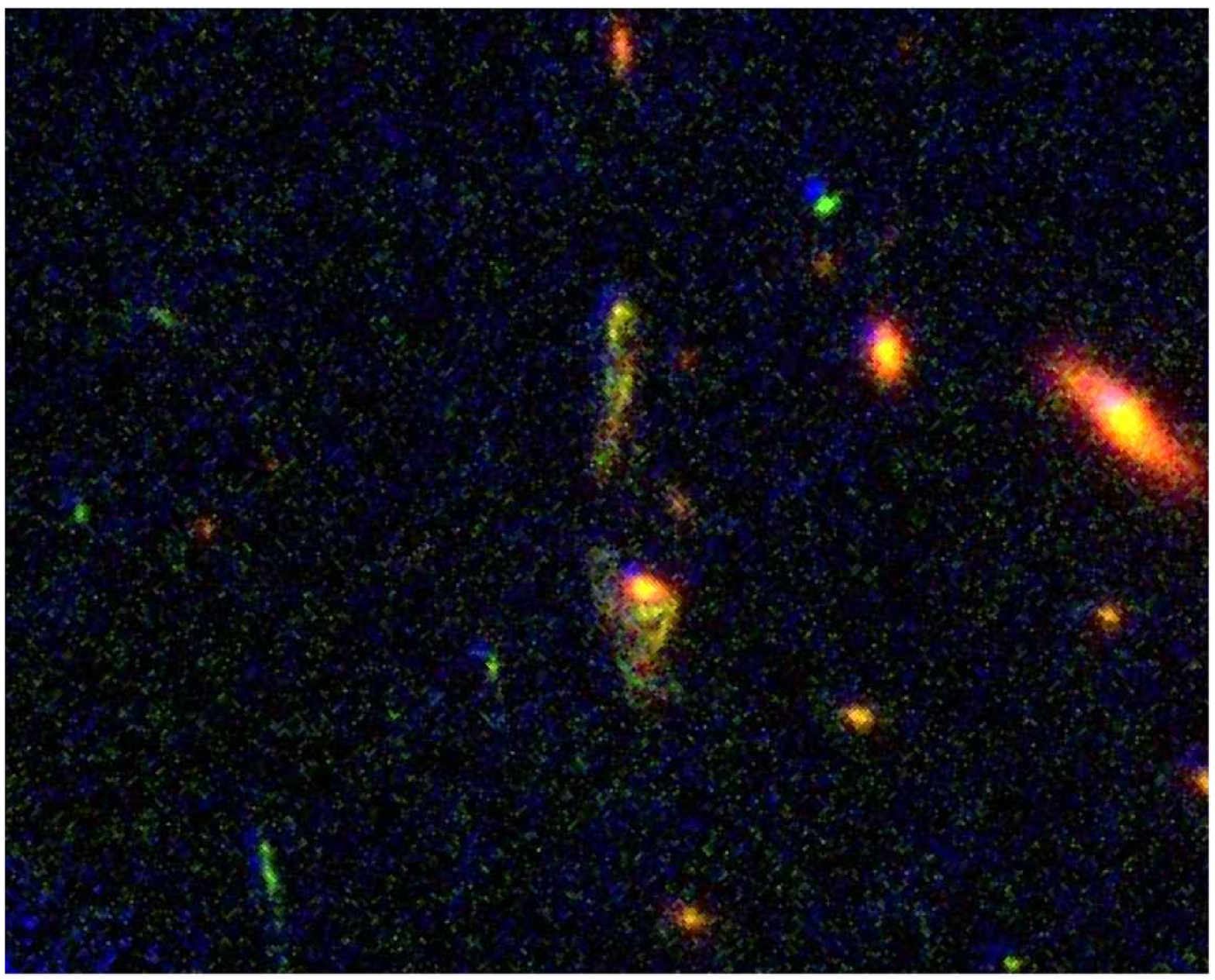} &
\includegraphics[width=0.33\textwidth]{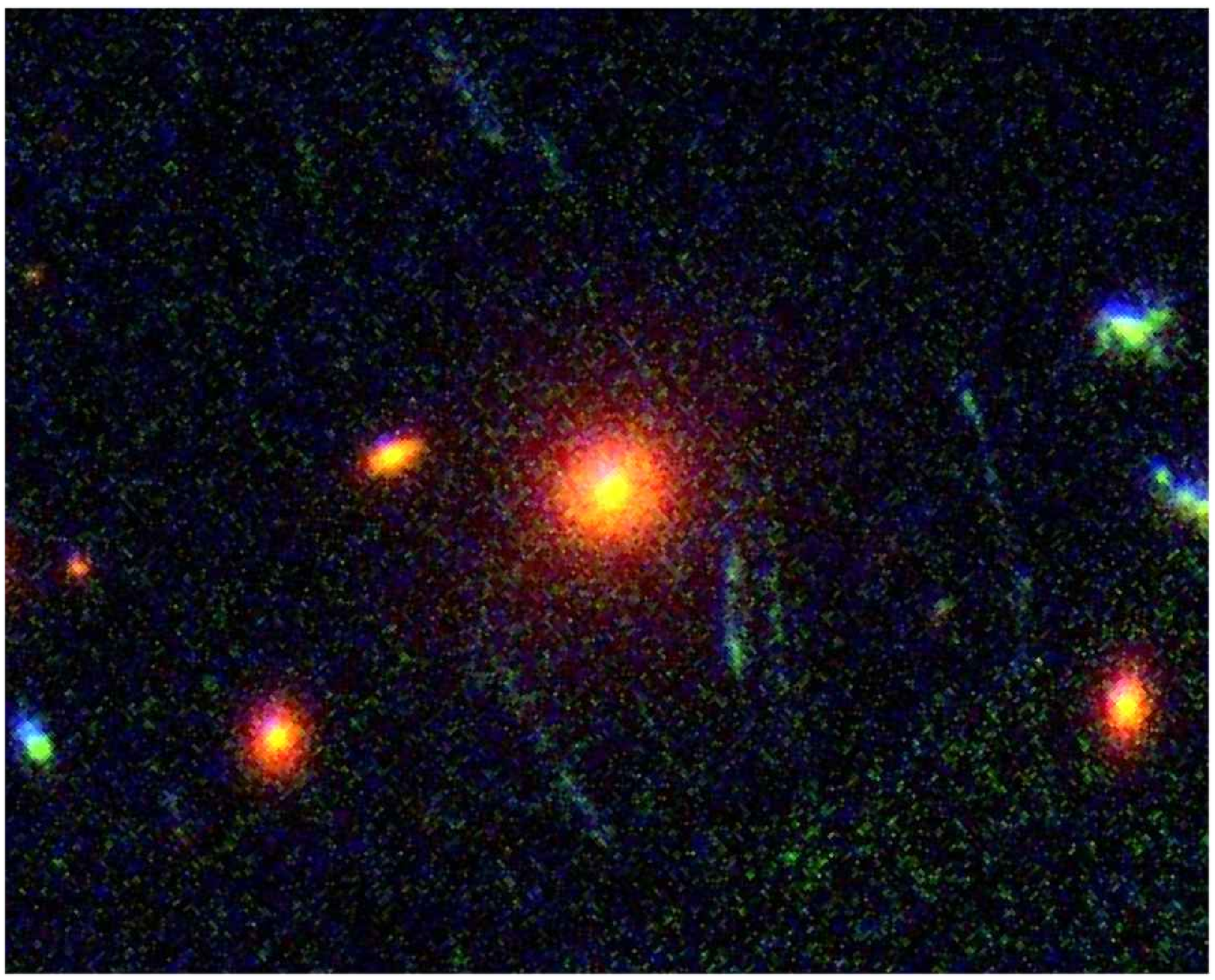} &
\includegraphics[width=0.33\textwidth]{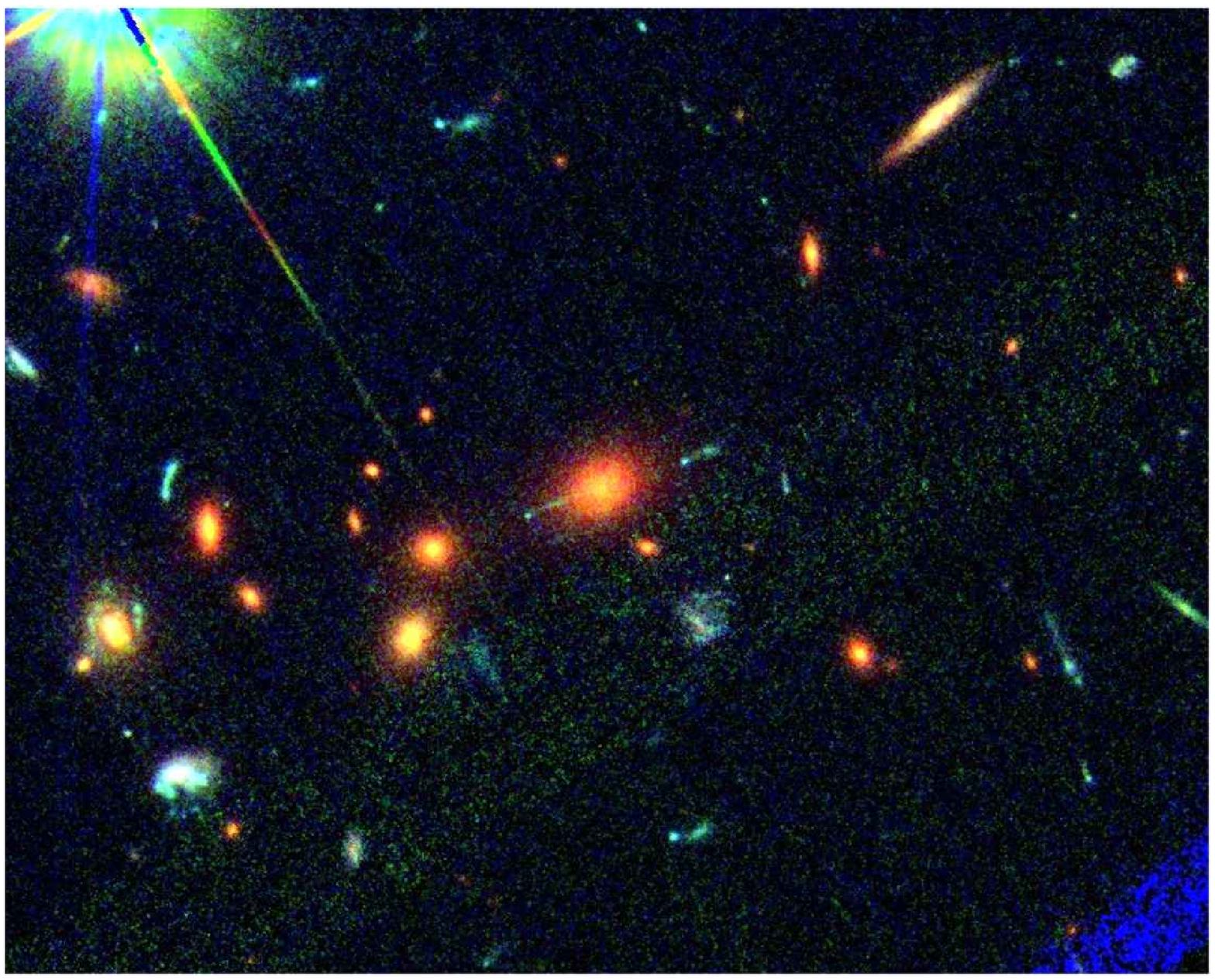} \\
C & D & AB\\
\end{tabular}
\end{center}
\caption{(top) The F555W-F814W color composite of the cluster
  \protect \bbullet. North is up and East is left. The field is
  $2.5^{\prime}\times 2.0^{\prime}$.  $15\arcsec \times
  15\arcsec$ cutouts (indicated in top panel) of C and D and $30\arcsec \times 30\arcsec$ of the multiply imaged
  systems AB from the F555W-F814W (middle row) and F450W-F555W-F814W (bottom row)
  color composite are also shown. The three BCGs are marked with circles, the gas peak position with a $+$ (with its size indicating the measured uncertainty, see \S\ref{sect:xray_analysis}).}
\label{fig:arcs}
\end{figure*}

Using the F450W/F555W/F814W HST and BVRIz' Subaru data, we have
searched for multiply-imaged, lensed background sources. The positions
of the multiple images used in the analysis are
listed in Table~\ref{tab:arcs} and shown in Fig.~\ref{fig:arcs}.

\begin{figure}[ht!]
\begin{center}
\includegraphics[width=0.5\textwidth]{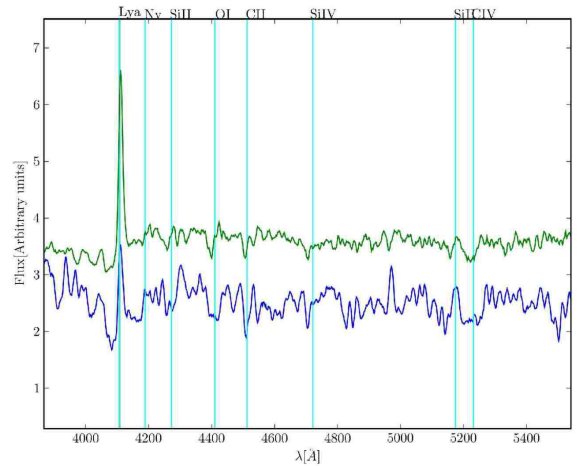}
\end{center}
\caption{LRIS spectrum of the images A and B (a single slit was placed over both images). 
Lines redshifted to $z=2.38$ are marked. Over-plotted (and offset for
clariy) in green (top) is the composite Ly-break galaxy spectrum from
\citet{shapley03}. The emission line is significantly detected
(6$\sigma$), no other significant features are seen in the spectrum.}
\label{fig:spec}
\end{figure}

We measure a spectroscopic redshift for system AB from the LRIS/KECK
data of $z=2.38$. This source has two distinct components (as seen from
Fig.~\ref{fig:arcs}; hence the name AB) and we placed slits over
both components each time. The spectrum of the northern of the three
images is shown in Fig.~\ref{fig:spec}, together with a composite
Ly-break galaxy template from \citet{shapley03}.  A Ly-$\alpha$
emission line is significantly detected (6$\sigma$). The other two
images had lower S/N spectra, but yield consistent results.  For
system C, we obtain a photometric redshift $z_{\rm C} =
1.0^{+0.5}_{-0.2}$ using the IRAF {\tt polyphot} task and the HyperZ
package \citep{bolzonella00}; this redshift is robust (i.e.\ there is
no other likely redshift solution) due to the presence of the
$4000$\AA\, break seen between the F555W and F814W bands. We have
placed a slit over the northern of the two images and we see a
tentative absorbtion feature at 7140\AA. If interpreted as \ion{Mg}{2}
at 2799{\AA} this gives a redshift of 1.55. However, the absorbtion
feature is detected at a very low significance and no other features
are seen. Therefore we do not consider this identification as reliable
and adopt the photometric redshift as our best estimate. Adopting the
tentative spectroscopic redshift instead would not change our results
significantly.

The photometric redshift of system D is less well determined. In
addition the spectra did not have high enough S/N to obtain any
spectroscopic redshift information. Using the photometric information
we estimate $z_{\rm D} = 2.8^{+0.4}_{-1.8}$.

The scarcity of reliably defined multiply imaged
systems and spectroscopic redshifts limits our ability to determine an
absolute calibration for the lensing mass map, due to the mass-sheet
degeneracy (i.e.\ the reconstruction suffers from a degeneracy of the
form $\kappa \to \kappa' = \lambda \kappa + (1 - \lambda)$, where
$\lambda$ is an arbitrary constant, see e.g.\ \citealp{bradac03b}).
This degeneracy can, in principle, be lifted when information from at
least two source-redshift planes is available (see e.g.\
\citealp{seitz97}; since $\kappa$ is redshift-dependent, data from two
redshifts can simultaneously constrain $\kappa$ and $\lambda$). In our
case, we have only one spectroscopically-determined source redshift
(around the NW peak). In principle, weak lensing data can provide a
second effective redshift plane, allowing us to tightly constrain the
mass, at least around the NW peak; in practice, however, the weak
lensing data are too noisy to completely break the mass-sheet
degeneracy (see e.g.\ \citealp{bradac03, limousin07}).  We might also,
in principle, constrain the redshifts of systems C and D around the SE
peak using the predictive power of the lens model. This requires the
adoption of a specific family of models; which includes assumptions
that lift the mass-sheet degeneracy, such as assuming zero mass at
large distances. Moreover the constraints in regions far from the
systems of known redshift tend to be weak.

\subsection{Weak lensing data}
\label{sec:weak_lensing}

Our weak lensing analysis of the HST/ACS data uses the RRG
\citep{rhodes00} galaxy shape measurement method, and the pipeline
developed for the HST COSMOS survey \citep{massey07}. This
has been tested and calibrated on simulated images containing a known
weak lensing signal by \citet{leauthaud07}.  The \bbullet\ and COSMOS
observations use the same filter and are contemporary, although the
COSMOS survey covered a much larger area of sky and provides a better
calibration of instrumental systematics, including the camera's point
spread function (PSF) and Charge Transfer Inefficiency (CTI). Thermal
variations within HST's orbit cause it to expand and contract by a few
microns \citep{rhodes07}. By matching the shapes of stars in the
\bbullet\ images to those in COSMOS fields \citep[which were in turn
compared to TinyTim models of the PSF, see][]{drizzle}, we determined
that the focal length of HST (the distance between its primary and
secondary mirrors) was $2\mu m$ shorter than nominal at the time of
ACS exposure. This is a typical value \citep{leauthaud07}. We created
a model of the PSF over the field of view and over all possible focus
positions by interpolating the shapes of stars in all images within
the COSMOS survey. Similarly, the PSF-corrected ellipticities of
galaxies in the \bbullet\ data were corrected for CTI trailing
(associated with radiation damage in the ACS detectors) by subtracting
the spurious ellipticity induced by this effect. This spurious
ellipticity was modelled from the observed shapes of galaxies within
the COSMOS survey as a function of object flux and position on the
detectors \citep{rhodes07}.

From the final catalog we selected background objects, excluding
cluster members using color information from the HST
(F450W/F555W/F814W) images and spectroscopic data. Galaxies in the
cluster red sequence are identified by two color criteria: $0.0 <
m_{\rm F555W} - m_{\rm F450W} < 1.4$ and $1.8 < m_{\rm F555W}- m_{\rm
F814W} < 3.0$.  We furthermore exclude objects brighter in F814W than the three
BCGs. We note that these cuts will not remove blue cluster members
(and some bluer or redder foregrounds), which will dilute the weak
lensing signal. However since the signal is mainly driven by the
strong lensing data (see also \citealp{bradac08}) this will not
bias our results significantly. Following \citet{bradac08}, where data
of similar depth were used, we assume the average redshift for galaxies
in the weak lensing catalog to be $z_{\rm WL} = 1.4$. Again the exact
value has little influence on the end results (as discussed in
\S\ref{sec:syst}) because it does not change the location of the peaks
and only marginally affects the overall scaling of the surface mass
density.

\begin{deluxetable}{rrrr}
\tablecolumns{4}
\tablewidth{0pc}
\tablecaption{The properties of the multiply-imaged systems.}
\tablehead{ \colhead{} &  \colhead{RA} & \colhead{Dec} & \colhead{$z$}}
\startdata
 & 6.365628 & -12.36976 &  \\ 
{A} & 6.363495 & -12.37609 & {2.38} \\ 
& 6.363393 & -12.37276 & \\
\cline{1-4}
 & 6.365353 & -12.36974 &  \\
{B} & 6.363261 & -12.37606 & {2.38} \\
 & 6.363169 & -12.37268 & \\
\cline{1-4}
 & 6.392082 & -12.38603 & \\
\raisebox{0.5ex}{C} & 6.391968 & -12.38757 & \raisebox{0.5ex}{$1.0^{+0.5}_{-0.2}$} \\
\cline{1-4}
 & 6.382898 & -12.38511 & \\
{D} & 6.384163 & -12.38253 &{$ 2.8^{+0.4}_{-1.8}$} \\
 & 6.382926 & -12.38472 & \\
\cline{1-4}
\enddata
\tablecomments{ 
The redshift of systems A and B were obtained using KECK spectroscopy (a
single slit was placed over both components each time), whereas the others have been
determined photometrically.}
\label{tab:arcs}
\end{deluxetable}

\subsection{Mass reconstruction of \protect\bbullet}
\label{sec:results}

Our combined strong and weak lensing analysis follows the method
described in \citet{bradac04a} and implemented on ACS data in
\citet{bradac06, bradac08}. We describe the cluster's
projected gravitational potential by a set of values on a regular grid
$\psi_k$, from which all observable quantities are evaluated by finite
differencing. For example, the scaled surface mass density $\kappa$ is
related to $\psi$ via the Poisson equation, $2\kappa = \nabla^2\psi$
(where the physical surface mass density is $\Sigma = \kappa \:
\Sigma_{\rm crit}$ and $\Sigma_{\rm crit}$ is a constant that depends
upon the angular diameter distances between the observer, the lens,
and the source). Similarly, the shear $\gamma = \gamma_1 + {\rm i}
\gamma_2$ and the deflection angle $\vec \alpha$ are derivatives of
the potential $\gamma_1 = 0.5(\psi_{,11}-\psi_{,22})$, $\gamma_2 =
\psi_{,12}$, and $\vec \alpha = \nabla \psi$.  The advantage of such
an approach is that we avoid strong assumptions regarding the shape
and profile of the potential, which is crucial when dealing with
merging clusters.

The strong and weak lensing data are combined in a $\chi^2$-fashion,
minimizing the $\chi^2$ difference between the data and the model by
searching for a solution to the nonlinear equation $\partial \chi^2 /
\partial \psi_k =0$. Hence we solve it
in an iterative fashion (keeping the non-linear terms fixed at each
iteration step). This requires an initial guess for the gravitational
potential; the systematic effects arising from these particular
choices are discussed below.

Figure~\ref{fig:kappa} shows our best-fit reconstruction of the
surface mass density, starting from an initial model consisting of two
isothermal spheres, each of $\sigma=1000\kms$. The first sphere was
centered between the two brightest galaxies in the SE cluster (BCG1
and BCG2), and the second on the BCG of the NW subcluster (BCG3, see
also Fig.~\ref{fig:arcs}). We also examined models in which the
initial velocity dispersion of the spheres was reduced to
$\sigma=700\kms$, and where the initial guess had zero mass: the
resulting reconstruction did not change significantly. Throughout the
paper we will quote the results with the initial model using
$\sigma=1000\kms$ as this model gives the best fit to the data. The other final solutions (encountered due to secondary minima in $\chi^2$ function) were however not significantly different.

The recovered distribution of mass in {\bbullet} shows two distinct
peaks, each centered approximately at the centers of the light
distributions (galaxies). The measured positions and error bars are
given in Table~\ref{tab:offset}. The quoted error bars reflect the
effects of exploring different initial conditions and other systematic
effects (like misidentification of multiply imaged systems), and the
statistical and systematic errors in the weak lensing data (see
below).  Table~\ref{tab:mass} lists the projected masses within
$300\mbox{ kpc}$ of each of the BCG1 and BCG3, and the mass within
$500\mbox{ kpc}$ of the X-ray peak.

\subsection{Systematic and statistical uncertainties}
\label{sec:syst}

\begin{figure*}[ht!]
\begin{center}
\includegraphics[width=0.7\textwidth]{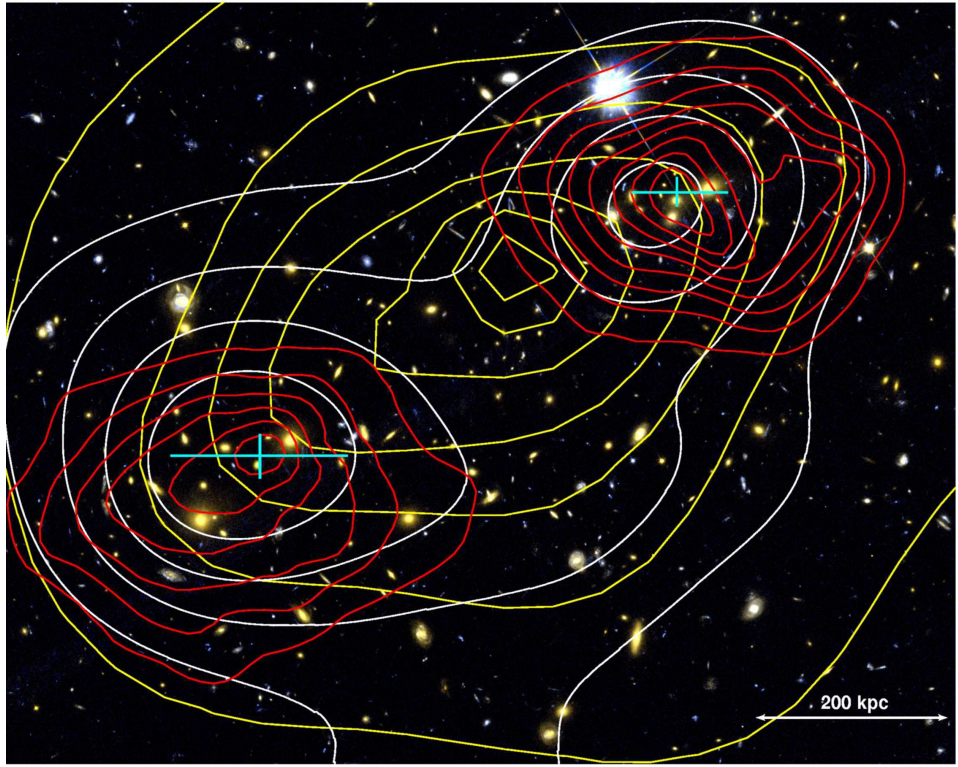}
\end{center}
\caption{The F555W-F814W color composite of the cluster
  \protect \bbullet. North is up and East is left. The field is
  $2.5^{\prime}\times 2.0^{\prime}$, which corresponds to $1000 \times
  800 \mbox{ kpc}^2$ at the redshift of the cluster.  Overlaid in {\it
  red contours} is the surface mass density $\kappa$ from the combined
  weak and strong lensing mass reconstruction. The contour levels are
  linearly spaced with $\Delta\kappa = 0.1$, starting at $\kappa=0.5$,
  for a fiducial source at a redshift of $z_{\rm s} \to \infty$. The
  X-ray brightness contours (also linearly spaced) are overlaid in
  {\it yellow} and the I-band light is overlaid in {\it white}. The
  measured peak position and errorbars for the total mass are given as a cyan cross (see
  Tab.~\ref{tab:offset}).}
\label{fig:kappa}
\end{figure*}

As in previous work, we pay special attention to possible systematic
effects when obtaining the surface mass density map. We have performed the 
following tests of robustness:
\begin{itemize}
\item {\bf Initial Conditions:} Instead of assuming a two component mass
model, we have examined the effects of starting with an ``ignorant''
model, i.e.\ assuming zero mass for the system. The final
reconstructed peak positions differ little from those shown in
Figure~\ref{fig:kappa} and are well within the quoted error bars. We obtained a better fit, however, when using two isothermal spheres as
initial conditions, hence the quoted numbers are from this 
reconstruction.
\item {\bf Multiply imaged system identification and redshifts:}
Misidentification of multiply imaged systems can, in principle, 
significantly change the
outcome of a reconstruction. We have examined the effects of
removing one system at a time, as well as removing them all. While
the absolute calibration of the mass distribution changes, the peak
positions are not significantly altered and these 
changes are added in quadrature to the quoted error bars.
\item {\bf Statistical errors on weak lensing measurements:} The
  observed ellipticities of galaxies provide a noisy estimate of the
  local reduced shear $g = \gamma / (1-\kappa)$. We
  bootstrap resample the measured ellipticities to estimate the uncertainties.
  These are added to the errors discussed above to obtain the final
  estimate of the uncertainty on the measured positions and masses of
  the two peaks.
\item {\bf Systematic error budget on weak lensing measurements:}
  We estimate a $\sim 6$\% uncertainty in the overall normalisation of
the shear measurements. There is an additional $\sim 15\%$ calibration
uncertainty from the assumption on the average redshift for galaxies
in the final catalog. These uncertainties only affect the mass
measurement (for which the systematic errors mentioned above dominate)
and do not add to the positional uncertainty.
\end{itemize}

\begin{deluxetable}{rlrr}
\tablecolumns{4} \tablewidth{0pc}
\tablecaption{Offsets (with 1$\sigma$ errorbars) of the mass peaks and gas peaks.}
\tablehead{ \colhead{} & \colhead{} & \colhead{$\Delta x ([\mbox{arcmin}])$} &  \colhead{$\Delta y ([\mbox{arcmin}])$}}
\startdata
 & total & $-0.70 \pm 0.25$ & $-0.43 \pm 0.06$ \phn \\
SE & gas& $0.0 \pm 0.08$ & $0.0 \pm 0.08$ \phn \\
 & galaxies & $-0.79\pm 0.20$ & $-0.50\pm 0.20$ \phn\\
\hline
 & total & $0.46 \pm 0.13$ & $0.20 \pm 0.04$ \phn \\
NW & gas & $0.0 \pm 0.08$ & $0.0 \pm 0.08$ \phn \\
 & galaxies &$0.39 \pm 0.20$ & $0.18 \pm 0.20$\phn 
\enddata
\tablecomments{ 
Offsets are calculated from (00:25:29.5, -12:22:36.6) which  corresponds to the peak in gas distribution (marked with a white $+$ in \protect{Fig.~\ref{fig:arcs}}).}
\label{tab:offset}
\end{deluxetable}

\begin{deluxetable}{rlr}
\tablecolumns{3} \tablewidth{0pc}
\tablecaption{{\it 2-D projected} enclosed mass within a radius of $300 \mbox{kpc}$ centered on BCG1 and BCG3 and within  $500 \mbox{kpc}$ centered on the gas peak.}
\tablehead{ \colhead{} & \colhead{} & \colhead{$M \; [10^{14}M_{\sun}$]}}
\startdata
SE - BCG1 & total & $2.5^{+1.0}_{-1.7}$\phn\\
 & galaxies &   $0.027 \pm 0.008$\phn\\
\hline
NW - BCG3 & total &  $2.6^{+0.5}_{-1.4}$\phn\\
 & galaxies & $0.019 \pm 0.006$\phn \\
\hline
Gas peak & total & $6.2 ^{+1.2}_{-4.0}$\phn\\
         & gas & $0.55 \pm 0.06$\phn\\
	 & galaxies & $0.05 \pm 0.01$
\enddata
\tablecomments{Total projected masses are calculated from strong and
  weak lensing reconstruction, gas mass from Chandra X-ray data, and
  stellar mass from F814W data.}
\label{tab:mass}
\end{deluxetable}

\section{X-ray data analysis}
\label{sect:xray_analysis}
Using the Chandra data, we have constructed an exposure-corrected,
0.5--7.0keV, 4 arcsec/pixel count image of the central 4x4 arcmin
region of the cluster. Two point sources in this region were removed,
and the CIAO \texttt{dmfilth} tool used to fill in the excised
regions by sampling from the distribution of surrounding pixels. The
image was then adaptively smoothed using the \texttt{csmooth} tool,
applying a sliding Gaussian kernel. The linearly-spaced contours
derived from this image are shown in yellow in Fig.~\ref{fig:kappa}.
The peak position in the adaptively smoothed X-ray contours is
00:25:29.5, -12:22:36.6. Although the precise position of this peak
depends in detail on the smoothing algorithm, we conservatively
estimate the accuracy to be better than 5 arcsec (i.e., the X-ray
brightness peak is constrained to lie within 5 arcsec of this
position). Significant X-ray emission from the cluster is detected out
to a radius $\sim$ 150 arcsec (1 Mpc).

The overall X-ray morphology of {\bbullet} exhibits 
characteristics typical of a large merger event: the cluster shows a
modest central surface brightness, high overall ellipticity and clear
shifting of the X-ray centroid as a function of radius. Fitting a King
model to the X-ray surface brightness profile, we measure a large core
radius of $\sim 60$ arcsec (400kpc) which is also typical for 
large, merging clusters (e.g.\ \citealp{allen98}).

We have carried out a spectral analysis of the Chandra data using the
XSPEC code (version 11.3.2; \citealp{arnaud96}). Background spectra
were extracted from appropriate source-free regions of the ACIS-I
chips.  The Galactic absorption was fixed at the value of $2.5 \times
10^{20}$ cm$^{-2}$ derived from HI surveys \citep{kalberla05}. We
use the MEKAL plasma emission code (\citealp{kaastra93};
incorporating the Fe-L calculations of \citealp{liedahl95}) and the photoelectric absorption models
of \citet{balucinska92}. The abundances of the elements were assumed
to vary with a common ratio, $Z$, with respect to Solar
values \citep{anders89}. Only data in the 0.6--7.0 keV energy range
were used.

\begin{figure}
\begin{center}
\includegraphics[width=0.35\textwidth,angle=270]{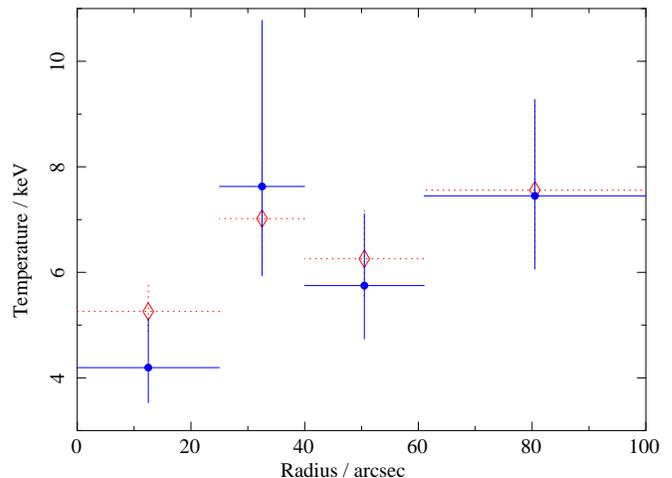}
\end{center}
\caption{Temperature profiles of the X-ray gas. Red diamonds show the
  projected values, blue circles the deprojected values (assuming
  spherical symmetry and using a \texttt{projct} model to account for
  geometric effects). Error bars are 1$\sigma$.}
\label{fig:Xtemp}
\end{figure}

\begin{figure}
\begin{center}
\includegraphics[width=0.38\textwidth,angle=270]{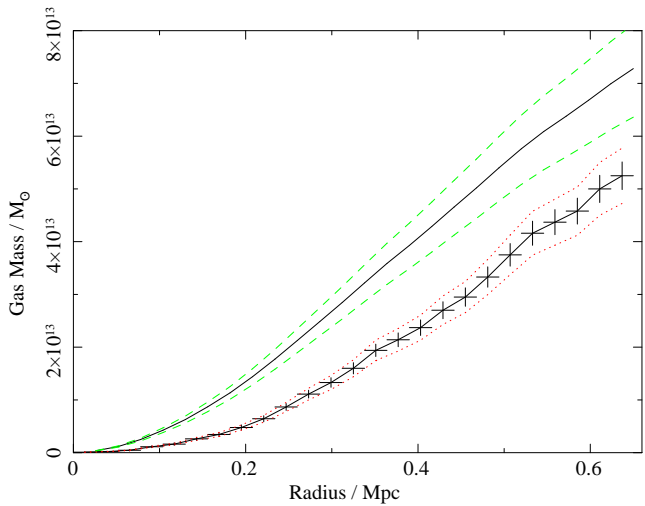}
\end{center}
\caption{Mass profiles of the X-ray gas. The lower curve shows the
  mass within a sphere, the upper curve the projected mass within a
  cylinder along the line of sight, assuming the cluster extends to
  1.5Mpc. The crosses show the statistical 1$\sigma$ errors
  associated with each point, as obtained from the method of \citet{allen08}. The red dotted lines show a (conservative) 10\%
  systematic error (e.g.\citealp{nagai07b}). The green dashed lines
  combine in quadrature the effect of a 10\% systematic error on the
  projected mass with the effect of varying the outermost cluster
  radius from 1.1 to 2.0Mpc.}
\label{fig:Xmass}
\end{figure}

Examining a circular region of radius $100$ arcsec (660 kpc) centered
on the X-ray peak, we measure a temperature of $6.26^{+0.50}_{-0.41}$
keV and a metallicity $Z=0.37 \pm 0.10$ solar. This spectrum contains
approximately 4300 net (background-subtracted) counts. Dividing this
region into four annuli with approximately equal net counts, we obtain
the (projected and deprojected) temperature profiles shown in
Fig.~\ref{fig:Xtemp}. The deprojected metallicity rises from a value
of 0.2 solar in the outer regions to $> 1.0$ within the central 25
arscec ($\sim$ 160 kpc).

\section{The gas and stellar mass content}
\label{sec:stellarmass}

The gas mass profile has been determined, under the assumption of
spherical symmetry, using the deprojection method described by  
\citet{allen08}. As discussed by e.g.\ \citet{nagai07b},
such an approach should provide an accurate measure of the gas mass,
with systematic errors of $<10\%$, even for clusters
experiencing major merger events. The deprojected gas mass profile,
with statistical and (conservative) systematic error bars, is shown as
the lower curve in Fig.~\ref{fig:Xmass}. Within a spherical radius of
500 kpc, we measure a gas mass of $3.6 \pm 0.4 \times 10^{13}$
M$_{\sun}$. The \textit{projected} gas mass within a cylinder along
the line of sight is shown in the upper curve; this result is
(slightly) sensitive to the outermost extent assumed for the cluster. 
The surface brightness profile is measurable out to 1.1Mpc, whereas 
a virial
radius for a cluster of this temperature and redshift is expected to 
be $\sim 1.5$Mpc. We
therefore consider values from 1.1--2Mpc for the outermost cluster radius.
Within the region plotted (for which X-ray temperature data are
available), the variation in projected mass due to our choice of outer
radius is less than 10\% at all radii. The error-range displayed in
Fig.~\ref{fig:Xmass} is formed from the sum in quadrature of this
uncertainty and a conservative 10\% systematic error. At 500kpc we measure
a projected mass $(5.5 \pm 0.6) \times 10^{13}$~M$_{\sun}$.

We adopt the following procedure to measure the stellar mass
distribution from the HST data. We identify cluster members based on
color-color diagrams, with the color cuts described in
\S\ref{sec:weak_lensing}, making use of the DEIMOS/KECK
spectroscopic redshifts when available. These cuts primarily select
red cluster members, which should be the dominant contributors to the
total luminosity. To calculate the stellar mass, we transform observed
$I$-band magnitudes (MAGAUTO,
\citealp{sextractor}) to K-band rest-frame luminosities, adopting an
absolute solar magnitude $M_{\rm K,\sun} = 3.28$, Galactic extinction
following \citet{ned1} and \citet{ned2} with $A_{\rm I}= 0.054$, and
K-correcting (we assume zero evolutionary correction) using an SED
template for an elliptical galaxy from \citet{coleman80} (extended in
the UV and IR regions with the GISSEL spectra,
\citealp{bruzual93}). The resulting K-correction is $2.464$. 

We note that our template is likely slightly redder than the cluster
members (possibly giving a larger stellar mass), although the
approximation is adequate for our purposes. The resulting luminosity
distribution (smoothed with a Gaussian kernel with FWHM $80\mbox{
kpc}$) is shown in Fig.~\ref{fig:kappa}. To convert luminosities to
stellar masses we follow \citet{drory04} and assume the
stellar-mass-to-light ratio in K-band to be $M_{*}/L_{\rm K}= 0.74\pm
0.30$ (see also \citealp{bell03}).

The enclosed, projected total mass within 500 kpc of the X-ray peak,
and the projected masses in X-ray emitting gas and stars (in galaxies)
are listed in Table~\ref{tab:mass}. The gas mass fraction of $9^{+7}_{-3}\%$ is consistent with the typical value of $\sim 11\%$ observed at
similar radii in other massive, X-ray luminous clusters
(e.g.\citealp{allen08}; note that the value of $11\%$ does not refer to the projected values, but the projection will make little
difference given the errorbars). The mass-to-light
ratio ($M/L_{\rm K} ( < 500\mbox{kpc}) = 130^{+40}_{-80}$) and stellar
mass fraction ($1.0^{0.7}_{0.4}\%$) within this aperture are also normal (see e.g.\
\citealp{kneib03,lin04}). Studying the inner regions of
the two subclusters, we measure mass-to-light ratios of $M/L_{\rm K} (
< 300\mbox{kpc}) = 70^{+40}_{-50}$ and $100^{+40}_{-60}$ for the SE
and NW peaks respectively, which are also consistent with values found
for other clusters (see e.g.\ \citealp{kneib03}). Both subclusters
have a stellar-to-total mass ratio of 1 per cent within $300\mbox{
kpc}$.

\section{Dissecting {\protect\bbullet} into dark matter and baryons}
\label{sec:bardm}
Figure~\ref{fig:kappa} shows the distributions of galaxies, X-ray
emitting gas and total mass in {\bbullet}. The figure clearly
demonstrates our main result: an obvious spatial offset is observed
between the peak of the X-ray emission and the twin peaks of the total
mass distribution. As in the case of {\bullet}, we see a good
alignment between the total mass and light distribution. The offsets
between the X-ray and total mass components are significant at
$>4\sigma$ for both the SE and NW clumps (see
Table~\ref{tab:offset}). We calculate the significance by assuming
that the errors on the total mass distribution follow Gaussian
distribution and those on the peak gas position a top-hat
function. This is conservative, since both strong and weak lensing
(separately) give the peak positions within the observed multiple
images, hence the lensing errors are also unlikely to be Gaussian.

Like the Bullet Cluster, the morphology and properties of {\bbullet}
are consistent with those of a high velocity merger event
approximately in the plane of the sky. Given a sound speed for the
X-ray emitting gas of $\sim 1300\kms$ and a collision velocity of
order $2000\kms$, the merger is likely to have been supersonic;
clearly it was violent enough to have separated the baryonic and dark
matter cores. The existing, short Chandra exposures are insufficient
to indentify and measure shock fronts in the cluster; a deeper Chandra
observation of {\bbullet} is required. Such a Chandra
observation would also permit detailed thermodynamic mapping of the
cluster \citep[e.g.][]{million08}.

In contrast to {\bullet}, neither of the subclusters in {\bbullet} contains
an X-ray bright, relatively cool, dense `bullet' of low-entropy gas. This is
not suprising; at low-to-intermediate redshifts only a fraction (tens of per
cent) of clusters contain such cores \citep[e.g.][]{peres98,bauer05}. Once
formed, these cores are expected to be relatively robust and, after the
merger event, eventually settle at the base of the newly-formed cluster
potential. Thus, whereas in $1-2$ Gyr time {\bbullet} may still not possess
a cool core, {\bullet} appears likely to evolve into a cool-core cluster.

\subsection{Limits on self-interaction cross-section and dark matter alternatives}
\label{sec:dmcs}

Following \citet{markevitch04} and based on the new data for {\bbullet},
we can make an order-of-magnitude estimate for the self-interaction
cross-section of dark matter. In the first case, 
the observed offset between the gas and
total mass component implies that the scattering depth of dark matter
particles $\tau = \Sigma\sigma/m$ during the collision cannot
be much greater than $1$. The surface mass density $\Sigma$ at the
position of both peaks averaged over a radius of $150\mbox{kpc}$ is
$\sim (0.25 \pm 0.10) \mbox{g}\:\mbox{cm}^{-2}$. Assuming that the two
subclusters experienced a head-on collision within this radius, we obtain
an estimate for the scattering cross-section of $\sigma/m < 4\cs$. The
assumption of spherical symmetry (we assume that the surface mass
density along the collision direction is similar to that along the
line-of-sight) appears reasonable in {\bbullet} with both subclusters
showing little ellipticity.

Secondly, the apparent survival of the dark matter content after the
recent merger event, implied by the mass-to-light ratios, confirms
that the dark matter needs to be
collisionless \citep{markevitch04}. Unfortunately current constraints
on the escape velocity and shock velocity are too weak to derive more
stringent limits on the cross-section. Better X-ray data, improved
lensing constraints and detailed hydrodynamical simulations are
required to significantly improve upon on these constraints.

Just like {\bullet}, {\bbullet} severely challenges modified gravities
theories such a MOND \citep[e.g.][]{milgrom83} and TeVeS
\citep{bekenstein04}. \citet{angus06} attempt to explain the weak
lensing data for {\bullet} using a MOND-like model with $2\mbox{ eV}$
neutrinos.  However, such neutrinos alone cannot explain the X-ray
properties of groups \citep{angus07}, and lie at the limit of the
allowed region from current neutrino experiments (Maintz/Troitsk;
\citealp[e.g.][]{weinheimer03}). Moreover, such as model cannot easily
explain the tight strong lensing constraints on the mass distributions
in the inner regions of {\bullet} and {\bbullet}, where multiple
images are observed.

\section{Conclusions and Outlook}
\label{sec:conclusions}

We have presented new data for the post-merging galaxy cluster
{\bbullet}. This system exhibits many similar properties to the
Bullet Cluster {\bullet}, although it does not contain a
low-entropy, high density hydrodynamical `bullet'. HST and Chandra
data have allowed us to derive a high-resolution total mass map, as
well as the X-ray gas mass and stellar mass distributions in the
cluster. Our main conclusions are as follows:
\begin{enumerate}
\item The total mass distribution in {\bbullet} shows two distinct
  peaks. Both are clearly offset from the main baryonic
  component (the hot gas) as traced by the X-ray emission, at $>4\sigma$
  significance (for each of the peaks). There is no significant offset
  between the peaks of the total mass distribution and the galaxies.
\item We have measured the projected, enclosed mass of the three main
mass constituents (dark matter, gas, and stars) within 500 kpc of the
X-ray center. The gas-to-total mass ratio ($9^{+7}_{-3}\%$) and the mass-to-light
ratio ($1.0^{+0.7}_{-0.4}\%$) are typical for a massive cluster.
\item The mass-to-light ratios $M/L_{\rm K}$ ($< 300\mbox{kpc}) =
  70^{+40}_{-50}$ and $100^{+40}_{-60}$ for the SE and NW peak, 
  and the stellar-to-total mass ratios ($1.0^{+0.7}_{-0.4}\%$) in these
  regions are consistent with values for other
  massive clusters, arguing that there was no significant
  loss of dark matter from either subcluster during the
  collision.
\item The majority of the mass is spatially coincident with the
  galaxies which implies, just as for the Bullet
  Cluster, that the cluster must be dominated by a relatively
  collisionless form of dark matter. We obtain an estimate for the
  self-interaction cross-section $\sigma/m < 4\cs$.
\end{enumerate}

{\bbullet} is a remarkable cluster with properties that allow us to
improve our understanding of the details of cluster formation and
evolution, and the nature of dark matter. Our results can be further improved with additional strong lensing information to
better constrain the normalization of the mass model and better X-ray
data to improve the determination of the merger speed, and hence
improve the limits on the dark matter cross-section. Discovering more
clusters like {\bbullet} and {\bullet} will also improve the
systematic uncertainties on dark matter self-interaction cross-section
inherent to the study of a small number of these objects.
\vspace{0.3cm}

%===============================================================================

\begin{acknowledgements} We would like to thank Cheng-Jiun Ma for the work done on KECK spectra and Matt Auger for sharing his KECK/LRIS pipeline with us.  We would also like to thank Phil Marshall and Eli
Rykoff for many useful discussions and Alice Shapley and Chun Ly for
providing us the spectral templates. Support for this work was
provided by NASA through grant numbers HST-GO-11100 from the Space
Telescope Science Institute, which is operated by AURA, Inc., under
NASA contract NAS 5-26555 and NNX08AD79G. MB acknowledges support from
NASA through Hubble Fellowship grant \#~HST-HF-01206.01 awarded by the
Space Telescope Science Institute. TT acknowledges support from the
NSF through CAREER award NSF-0642621, by the Sloan Foundation through
a Sloan Research Fellowship, and by the Packard Foundation through a
Packard Fellowship. RM acknowledges support from STFC through Advanced
Fellowship \#~PP/E006450/1. SWA and RGM acknowledge support from the
U.S. Department of Energy under contract number DE-AC02-76SF00515.
This research has made use of data obtained from the Chandra Data
Archive and software provided by the Chandra X-ray Center (CXC). Some
of the data presented herein were obtained at the W.M. Keck
Observatory, which is operated as a scientific partnership among the
California Institute of Technology, the University of California and
the National Aeronautics and Space Administration. The Observatory was
made possible by the generous financial support of the W.M.\ Keck
Foundation. \end{acknowledgements}

\end{document}